\documentclass{article}

\usepackage{amsmath,amssymb}
\usepackage{dsfont}
\usepackage{bm}
\usepackage{tikz}
\usetikzlibrary{calc,matrix,arrows,decorations.pathmorphing,decorations.markings}

\renewcommand{\vec}[1]{{\ensuremath{\mathbf{\bm{#1}} }}}
\newcommand{\Dif}{\text{d}}
\newcommand{\braket}[3]{\ensuremath{\left< #1 \vphantom{#3} \right| #2 \left| #3 \vphantom{#1} \right>}}
\renewcommand{\i}{\text{i}}
\newcommand{\tr}{\, \text{tr}\,}

\title{Evolution and Dynamics of Cusped Light-Like Wilson Loops in Loop Space}

\author{F.F. Van der Veken$^\dag$, I.O. Cherednikov$^\dag{}^\ddagger$ and T. Mertens$^\dag$ \\
	${}^\dag\!\!\!\!$ \emph{ Departement Fysica, Universiteit Antwerpen, Belgium} \\
	${}^\ddagger\!\!$ \emph{BLTP JINR, Dubna, Russia}
}

\begin{document}

\maketitle

\begin{abstract}
We discuss the possible relation between the singular structure of TMDs on the light-cone and the geometrical behaviour of rectangular Wilson loops.
\end{abstract}

\section{Introduction}
Transverse momentum dependent parton density functions (or TMDs for short) are known to have a more complex singularity structure than collinear parton density functions. Common singularities like ultraviolet poles can be removed by general methods like standard renormalisation using the $R$-operation. In the case of  a light-like TMD however, where at least one of its segments is on the light-cone, it is not entirely clear whether standard renormalisation remains a sufficient technique, due to the emergence of extra overlapping divergencies. A standard TMD can be defined as \cite{TMDdef}:
\begin{equation}
	f(x,\vec{k}_\perp) = 
		\frac{1}{2}\int \frac{\Dif z^-\Dif^2 \vec{z}_\perp}{2\pi(2\pi)^2} \; e^{ik\cdot z}
		\braket{P,S}{\bar{\psi}(z) \, U^\dag (z;\infty) \gamma^+ U(\infty;0) \, \psi(0)}{P,S} \Big|_{z^+=0}
\end{equation}
where the Wilson lines are split into their longitudinal and transversal parts:
\begin{align}
	U(\infty, 0) &=
		U(\infty^-,\vec{\infty}_\perp ; \infty^-, \vec{0}_\perp) U(\infty^-,\vec{0}_\perp ; 0^-, \vec{0}_\perp)\\
	&=
		\mathcal{P} e^{-\i g \int_{0}^{\infty} \Dif z_\perp \; A_\perp(\infty^-, \vec{z}_\perp)}
		\mathcal{P} e^{-\i g \int_{0}^{\infty} \Dif z^- \; A^+(z^-, \vec{0}_\perp)}
\end{align}
When on light-cone, this TMD will posses extra divergencies proportional to $\frac{1}{\epsilon^2}$ (when using dimensional regularisation). These will give the only contribution to the evolution equations, governed by the cusp anomalous dimension \cite{cad1,cad2}
\begin{equation}
\label{eq:cad}
	\Gamma_{\text{cusp}} = \frac{\alpha_s \, C_F}{\pi} \left( \chi \coth \chi - 1 \right)
	\quad \stackrel{\text{on-LC}}{\longrightarrow} \quad
	\frac{\alpha_s \, C_F}{\pi}
\end{equation}
where $\chi$ is the cusp angle (in literature sometimes referred to as a `hidden cusp') which goes to infinity in the light-cone limit. In the next sections, we will show that a specific type of Wilson loop, namely rectangular loops with light-like segments on the null-plane, has its singularity structure analogous to on-LC TMDs, which feeds the idea that there might exist a duality between those two objects.

\section{Wilson Loops as Elementary Objects in Loop Space}
A general Wilson loop is defined as
\begin{equation}\label{eq:wloop}
	\mathcal{W}[C] =
		\frac{1}{N_c}\tr \braket{0}{\mathcal{P} e^{\i g \oint_C \Dif z^\mu A^a_\mu (z) t_a }}{0}
\end{equation}
where $C$ is any closed path and $A_\mu$ is taken in the fundamental representation. This loop is a pure phase, traced over Dirac indices and evaluated in the ground state, transforming coordinate dependence into path dependence. As is known (see \cite{QCDrecast1,QCDrecast2}), Wilson loops can be used as elementary objects to completely recast QCD in loop space. To achieve this, the definition of a Wilson loop needs to be extended to make it dependent on multiple contours:
\begin{align}
	\mathcal{W}_n(C_1,\ldots, C_n) &=
		\braket{0}{\Phi(C_1) \ldots \Phi(C_n)}{0} &      
	\Phi (C) &=
		\frac{1}{N_c}\tr \, \mathcal{P} e^{\i g \oint_{C} \Dif z^\mu A_\mu (z) }
\end{align}
All gauge kinematics are encoded in a $\mathcal{W}_1$ loop, and all gauge dynamics are governed by a set of geometrical evolution equations, the Makeenko-Migdal equations \cite{MMeqs}:
\begin{equation}
	\partial^\nu \frac{\delta}{\delta \sigma_{\mu\nu}(x)} \mathcal{W}_1(C) =
		g^2 N_c \oint_C \Dif z^\mu \delta^{(4)}\left(x-z\right) \mathcal{W}_2(C_{xz} \, C_{zx})
\end{equation}
where two (geometrical) operations are introduced, namely the path derivative $\partial_\mu$ and the area derivative $\frac{\delta}{\delta\sigma_{\mu\nu}(x)}$ \cite{MMeqs}. 
Although the Makeenko-Migdal equations provide an elegant method to describe the evolution of a generalised Wilson loop solely in function of its path, they have their limitations. For starters, they are not closed since the evolution of $\mathcal{W}_1$ depends on $\mathcal{W}_2$. Formally, this limitation is superfluous in the large $N_c$ limit since then we can make use of the 't Hooft factorisation property $\mathcal{W}_2 (C_1,C_2)\approx \mathcal{W}_1(C_1)\mathcal{W}_1(C_2)$ \cite{MMeqs}, making the MM equations closed. The remaining limitations of the MM equations are more severe. For one, the evolution equations are derived by applying the Schwinger-Dyson methodology on the Mandelstam formula
\begin{equation}
	\frac{\delta}{\delta \sigma_{\mu\nu} (x)} \Phi(C) =
		\i g \tr \! \left\{ F^{\mu\nu} \Phi(C_x) \right\}
\end{equation}
and using the Stokes' theorem. These might, as well as the area derivative, not be well-defined for all types of paths. In particular, all contours containing one or more cusps might induce some problematic behaviour, as it is (at least) not straightforward to define \emph{continuous} area differentiation inside a cusp, nor it is to continuously deform a contour in a general topology \cite{ourpaper}. This is somewhat bothersome, as most interesting dynamics lies in contours with cusps.

\section{Evolution of Rectangular Wilson Loops}
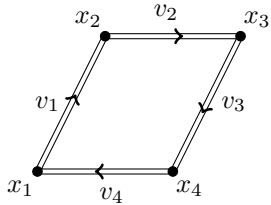
\begin{figure}[h!]
	\center
	\label{fig:rect}
	\begin{tikzpicture}[scale=0.9]
		\draw[double,double distance=1.5pt] (0,0) -- (1,2) -- (3,2) -- (2,0) -- cycle;
		\path[very thick, postaction={decorate}, decoration={markings,mark=between positions 0.15 and 1 step .25 with {\arrow{>}}}] (0,0) -- (1,2) -- (3,2) -- (2,0) -- cycle;
		\node at (-0.25,-0.25) {$x_1$};
		\node at (0.75,2.25) {$x_2$};
		\node at (3.25,2.25) {$x_3$};
		\node at (2.25,-0.25) {$x_4$};
		\node at (0.15,1) {$v_1$};
		\node at (1.9,2.35) {$v_2$};
		\node at (2.9,1) {$v_3$};
		\node at (1.1,-0.35) {$v_4$};
		\filldraw (0,0) circle(2pt);
		\filldraw (1,2) circle(2pt);
		\filldraw (3,2) circle(2pt);
		\filldraw (2,0) circle(2pt);
	\end{tikzpicture}
	\caption{Parametrisation of a rectangular Wilson loop in coordinate space.}
\end{figure}
Now we turn to a specific type of path, namely a rectangular path with light-like segments ($v_i^2 = 0$) on the null-plane ($\vec{x}_\perp=0$), as depicted in Figure (\ref{fig:rect}).
To investigate its singularity structure, we evaluate the loop \eqref{eq:wloop} at one loop in coordinate space \cite{wexpansion}:
\begin{equation}
	\mathcal{W}_{\text{L.O.}} =
		1 -\frac{\alpha_s C_F}{\pi} \left(2\pi \mu^2 \right)^\epsilon \Gamma (1-\epsilon) \left[
			\frac{1}{\epsilon^2} \left(-\frac{s}{2} \right)^\epsilon + \frac{1}{\epsilon^2} \left(-\frac{t}{2} \right)^\epsilon
			-\frac{1}{2}\ln^2\frac{s}{t}
		\right]
\end{equation}
where $s$ and $t$ are the Mandelstam energy/rapidity variables (note the positive sign in $t$):
\begin{equation}
	s = \left( v_1 + v_2\right)^2 \qquad\qquad t = \left( v_2 + v_3\right)^2 \qquad\qquad v_i = x_i - x_{i+1}
\end{equation}
Note the $\frac{1}{\epsilon^2}$ poles, which are the overlapping divergencies that stem from the light-like behaviour of the contour segments. The fact that they appear already at leading order renders this kind of Wilson loop non-renormalisable (at least not using the standard $R$-operation). The most straightforward way to manage them is by deriving an evolution equation for the loop. This is done by double differentiation (after rescaling $\bar{s} = \pi e^{\gamma_E} \mu^2 s$):
\begin{equation}\label{eq:evolution}
	\frac{\Dif}{\Dif \ln \mu}\frac{\Dif}{\Dif \ln s} \mathcal{W}_{\text{L.O}} =
		-2\frac{\alpha_s C_F}{\pi} = -2 \Gamma_{\text{cusp}}
\end{equation}
where we recognise the cusp anomalous dimension in the light-cone limit from \eqref{eq:cad}. Thus, as anticipated in the TMD case, the only contribution to the evolution equations stems from the overlapping divergencies. Their concurrent appearance in the on-LC TMD case and in the case of an on-LC rectangular Wilson loop again hints to the existence of a duality between both.

\section{Geometrical Behaviour and Relation to TMDs}
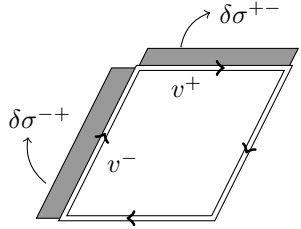
\begin{figure}[h!]
	\center
	\begin{tikzpicture}
		\filldraw[fill=black!40] (1,1) -- (1.13,1.26) --  (3.13, 1.26) -- (3,1) -- cycle;
		\filldraw[fill=black!40] (0.65,1) -- (1,1) --  (0, -1) -- (-0.35,-1) -- cycle;
		\begin{scope}[shift={(0,-1)}]
			\draw[double,double distance=1.5pt] (0,0) -- (1,2) -- (3,2) -- (2,0) -- cycle;
			\path[very thick, postaction={decorate}, decoration={markings,mark=between positions 0.15 and 1 step .25 with {\arrow{>}}}] (0,0) -- (1,2) -- (3,2) -- (2,0) -- cycle;
		\end{scope}
		\node[anchor=base] (up) at (2.5,1.6) {$\delta\sigma^{+-}$};
		\node at (1.5,1.15){}
			edge[->, bend left] (up);
		\node[anchor=base] (right) at (-0.4,0.2) {$\;\;\delta\sigma^{-+}$};
		\node at (-0.1,-0.65){}
			edge[->, bend left] (right);
		\node at (1.6 5,0.75) {$v^+$};
		\node at (0.8,-0.25) {$v^-$};
	\end{tikzpicture}
	\caption{Angle-conserving deformations of a light-like rectangular loop on the null-plane.}
	\label{fig:angdefor}
\end{figure}
In an attempt to combine the geometrical approach of the Makeenko-Migdal method with the evolution equations at leading order just derived, we investigate area differentiation on rectangular light-like loops on the null-plane, rendering the area differentials well-defined (see Figure (\ref{fig:angdefor})) \cite{ourpaper}. This gives
$\delta\sigma^{+-} =
	\oint \Dif x^- x^+ = v^+\delta v^-$ and
$\delta\sigma^{-+} =
	\oint \Dif x^+ x^- = v^-\delta v^+
$. 
Next we introduce the area variable $\Sigma$:
\begin{equation}
	\Sigma \equiv 
		v^-\cdot v^+ = \frac{1}{2} s \qquad\qquad
	\frac{\delta}{\delta\ln\Sigma} = 
		\sigma_{\mu\nu}\frac{\delta}{\delta\sigma_{\mu\nu}}
\end{equation}
Replacing $s$ by $\Sigma$ in equation (\ref{eq:evolution}) gives $-4\Gamma_{\text{cusp}}$.
Motivated by this, we conjecture a general evolution equation for light-like retangular Wilson loops:
\begin{equation}\label{eq:result}
\frac{\Dif}{\Dif\ln\mu} \left[\sigma_{\mu\nu}\frac{\delta}{\delta \sigma_{\mu\nu}} \ln \mathcal{W} \right] =
						-\sum_i \Gamma_{\text{cusp}}
\end{equation}
Besides for light-like rectangular Wilson loops, equation (\ref{eq:result}) is expected to be valid for light-like TMDs, as they posses the same singularity structure. The area variable then gets replaced by the rapidity variable. This gives
\begin{equation}
	\frac{\Dif}{\Dif \ln \mu}\frac{\Dif}{\Dif \ln \theta} f(x,\vec{k}_{\perp}) = 2\Gamma_{\text{cusp}}
\end{equation}
The minus disappeared because $\theta\sim \Sigma^{-1}$ ($\theta = \frac{\eta}{p\cdot v^-}$ and $p\sim v^+$, so $\theta \sim (v^+v^-)^{-1}$), and there is a factor 2 since we haven two (hidden) cusps. This result is very similar to the Collins-Soper evolution equations for off-LC TMDs.


\end{document}